# Einstein and the conservation of energy-momentum in general relativity

Galina Weinstein

10/10/2013

Abstract: the main purpose of the present paper is to show that a correction of one mistake was crucial for Einstein's pathway to the first version of the 1915 general theory of relativity, but also might have played a role in obtaining the final version of Einstein's 1915 field equations. In 1914 Einstein wrote the equations for conservation of energy-momentum for matter, and established a connection between these equations and the components of the gravitational field. He showed that a material point in gravitational fields moves on a geodesic line in space-time, the equation of which is written in terms of the Christoffel symbols. By November 4, 1915, Einstein found it advantageous to use for the components of the gravitational field, not the previous equation, but the Christoffel symbols. He corrected the 1914 equations of conservation of energy-momentum for matter. Einstein had already basically possessed the field equations in 1912 together with his mathematician friend Marcel Grossman, but because he had not recognized the formal importance of the Christoffel symbols as the components of the gravitational field, he could "not obtain a clear overview". Finally, considering the energy-momentum conservation equations for matter, an important similarity between equations suggests that, this equation could have assisted Einstein in obtaining the final form of the field equations (the November 25, 1915 ones) that were generally covariant.

In the 1914 paper, "The Formal Foundation of the General Theory of Relativity", Einstein showed that a material point in gravitational fields moves on a geodesic line in a four-dimensional continuum and satisfies the equation:[1]

$$(1) \sum_{\mu} g_{\sigma\mu} \frac{d^2 x_\mu}{ds^2} + \sum_{\mu\nu} \begin{bmatrix} \mu\nu \\ \sigma \end{bmatrix} \frac{dx_\mu}{ds} \frac{dx_\nu}{ds} = 0$$

so that,

$$(2) \begin{bmatrix} \mu\nu \\ \sigma \end{bmatrix} = \frac{1}{2}\left(\frac{\partial g_{\mu\sigma}}{\partial x_\nu} + \frac{\partial g_{\nu\sigma}}{\partial x_\mu} - \frac{\partial g_{\mu\nu}}{\partial x_\sigma}\right)$$

---

[1] $g_{\mu\nu}$ determine the gravitational field. For further details see my paper: Weinstein, Galina, "FROM THE BERLIN 'ENTWURF' FIELD EQUATIONS TO THE EINSTEIN TENSOR I: October 1914 until Beginning of November 1915", ArXiv: 1201.5352v1 [physics.hist-ph], 25 January, 2012, pp. 19-21; Einstein, Albert, "Die formale Grundlage der allgemeinen Relativitätstheorie", *Königlich Preußische Akademie der Wissenschaften* (Berlin). *Sitzungsberichte*, 1914, pp. 1030-1085; p. 1046.

are the Christoffel symbols of the first kind. By the Christoffel symbols of the second kind, (2) is,

$$(3) \begin{Bmatrix} \mu\nu \\ \tau \end{Bmatrix} = \sum_{\sigma} g^{\sigma\tau} \begin{bmatrix} \mu\nu \\ \sigma \end{bmatrix}$$

and then equation (1) becomes:

$$(4) \frac{d^2 x_\tau}{ds^2} + \sum_{\mu\nu} \begin{Bmatrix} \mu\nu \\ \tau \end{Bmatrix} \frac{dx_\mu}{ds} \frac{dx_\nu}{ds} = 0.$$

Einstein derived the "energy-momentum theorem… concerning 'material processes'." In Einstein's 1914 theory the "energy tensor" $T_{\sigma\nu}$ is a symmetric tensor of rank two. $T_{\sigma\nu}$ is represented by a mixed volume tensor $\mathfrak{T}_\sigma^\nu$. The generally covariant equations that represent energy-momentum conservation for matter in Einstein's notation are:

$$(5) \sum_{\nu} \frac{\partial \mathfrak{T}_\sigma^\nu}{\partial x_\nu} = \frac{1}{2} \sum_{\mu\tau\nu} g^{\tau\mu} \frac{\partial g_{\mu\nu}}{\partial x_\sigma} \mathfrak{T}_\tau^\nu + \mathfrak{K}_\sigma,$$

where,

$\mathfrak{K}_\sigma$ is a covariant volume four-vector, which is the representation of a four vector $K_\sigma$,

the three components of which are force vector per unit volume externally acting upon the system, and one component is the energy per unit volume and unit time supplied to the system.[2]

Einstein wrote for the effect of the gravitational field on matter, the "components of the gravitational field" (which entered into the 1914 field equations):

$$(6) \Gamma_{\nu\sigma}^\tau = \frac{1}{2} \sum_{\mu} g^{\tau\mu} \frac{\partial g_{\mu\nu}}{\partial x_\sigma}$$

In Einstein's 1914 theory, there was clear distinction between volume tensors ("V-tensors") and tensors.[3]

---

[2] See my paper: Weinstein, 2012, pp. 22-23.
[3] Einstein, 1914, pp. 1054-1056; p. 1058.

By November 4, 1915 Einstein realized that when only substitutions of determinant 1 are permitted volume tensors are superfluous. Einstein thus rewrote equation (5) in terms of ordinary tensors:[4]

$$(7) \quad \sum_{\nu} \frac{\partial T_\sigma^\nu}{\partial x_\nu} = \frac{1}{2} \sum_{\mu\tau\nu} g^{\tau\mu} \frac{\partial g_{\mu\nu}}{\partial x_\sigma} T_\tau^\nu + K_\sigma$$

$K_\sigma$ vanishes when $T_\sigma^\nu$ denotes the energy tensor of all "material" processes. Equation (7) for the total energy tensor of matter reads:

$$(8) \quad \sum_{\lambda} \frac{\partial T_\sigma^\lambda}{\partial x_\lambda} = -\frac{1}{2} \sum_{\mu\nu} \frac{\partial g^{\mu\nu}}{\partial x_\sigma} T_{\mu\nu}.$$

Einstein then explained, "This equation of conservation led me in the past to view the quantities [(6)] as the natural expression of the components of the gravitational field, even though in view of the formulas of the absolute differential calculus, it is better to introduce the Christoffel symbols […] instead of these quantities."[5]

Then when Einstein finally found the field equations that were generally covariant, he quickly saw that what he needed for general relativity which he was struggling to develop was the following, as he explained to Arnold Sommerfeld: "The key to this solution was my realization that not [(6)] but the related Christoffel's symbols are to be regarded as the natural expression of the gravitational field 'components'." And Einstein told Hendryk Antoon Lorentz: "I had already basically possessed the current equations 3 years ago together with Grossman, who had brought my attention to the Riemann tensor. But because I had not recognized the formal importance of the { } terms [Christoffel symbols], I could not obtain a clear overview and […] Then I fell into the jungle!"[6]

In 1914, Einstein substituted an equation[7] and this turned equations (2) and (3) to: [8]

---

[4] Einstein, Albert (1915a), "Zur allgemeinen Relativitätstheorie", *Königlich Preußische, Akademie der Wissenschaften* (Berlin). *Sitzungsberichte*, 1915, pp. 778-786; p.782.
[5] Einstein, 1915a, pp. 782-783.
[6] Einstein to Sommerfeld, November 28, 1915, *The Collected Papers of Albert Einstein. Vol. 8: The Berlin Years: Correspondence, 1914–1918* (*CPAE*, Vol. 8), Schulmann, Robert, Kox, A.J., Janssen, Michel, Illy, Jószef (eds.), Princeton: Princeton University Press, 2002, Doc. 153; Einstein to Lorentz, January, 1 1916, *CPAE*, Vol. 8, Doc. 177.
[7] Differentiating the determinant $|g_{\mu\nu}| = g$ with respect to $x_\alpha$, Einstein obtained the following equation: $\frac{1}{g}\frac{\partial g}{\partial x_\alpha} = \sum_{\mu\nu} \frac{\partial g_{\mu\nu}}{\partial x_\alpha} g^{\mu\nu}$.
[8] Einstein, 1914, p. 1051.

$$(9) \sum_\tau \begin{Bmatrix} \mu\nu \\ \tau \end{Bmatrix} = \sum_\tau \begin{Bmatrix} \nu\mu \\ \tau \end{Bmatrix} = \frac{1}{2} \sum_{\tau\alpha} g^{\tau\alpha} \frac{\partial g_{\tau\alpha}}{\partial x_\mu}$$

In the first general relativity paper of November 4, 1915, "On the General Theory of Relativity", from equations (2), (3), and (9) Einstein was led to the following new expression for the components of the gravitational field:

$$(10)\ \Gamma^\sigma_{\mu\nu} = -\begin{Bmatrix}\mu\nu \\ \sigma\end{Bmatrix} = -\sum_\alpha g^{\sigma\alpha}\begin{Bmatrix}\mu\nu \\ \sigma\end{Bmatrix} = -\frac{1}{2}\sum_\alpha g^{\sigma\alpha}\left(\frac{\partial g_{\mu\alpha}}{\partial x_\nu} + \frac{\partial g_{\nu\alpha}}{\partial x_\mu} - \frac{\partial g_{\mu\nu}}{\partial x_\alpha}\right).$$

With (10) the equations (4) of the geodesic line take the form:

$$(11)\ \frac{d^2 x_\tau}{ds^2} = \sum_{\mu\nu} \Gamma^\tau_{\mu\nu} \frac{dx_\mu}{ds}\frac{dx_\nu}{ds},$$

and equation (7) or (8) is given by:

$$(12)\ \sum_\lambda \frac{\partial T^\lambda_\sigma}{\partial x_\lambda} = -\sum_{\mu\nu} \Gamma^\mu_{\sigma\nu} T^\nu_\mu.$$

Einstein then proceeded to the field equations. Once having the components of the gravitational field, Einstein was able to write the new 1915 field equations of the general theory of relativity. In gravitation Einstein was interested in the Ricci tensor. Contracting the Riemann tensor results in the Ricci tensor: [9]

(13) $G_{im} = R_{im} + S_{im}$.

This division of equation (13) was already implicitly obtained by Einstein in 1912. Under the constraint to transformations with determinant equal to 1, $G_{im}$, $R_{im}$ and $S_{im}$ were all tensors. Einstein noted that the tensor $R_{im}$ was of utmost importance for gravitation. It replaced Einstein's problematic 1914 faulty gravitational tensor within a limited-covariant theory.

Einstein considered the tensor $R_{im}$, and wrote the general form of the field equations:

(14) $R_{\mu\nu} = -\kappa T_{\mu\nu}$.

---

[9] Note 5.

These equations are covariant with respect to arbitrary transformations of a determinant equal to 1. Hence, (14) were not yet fully generally covariant. (14) establish a connection between $R_{\mu\nu}$ (which includes the metric tensor and its derivatives) and $T_{\mu\nu}$ (the stress-energy tensor). The equation is non-linear because of $\Gamma^{\sigma}_{\mu\nu}$, which are defined by (10).[10]

In the final 1915 general relativity paper, presented on November 25, 1915, Einstein first wrote the field equations in the following form:[11]

$$(15)\ G_{im} = -\kappa \left( T_{im} - \frac{1}{2} g_{im} T \right), \quad \text{with:}$$

$$(16)\ \sum_{\rho\sigma} g^{\rho\sigma} T_{\rho\sigma} = \sum_{\sigma} T^{\sigma}_{\sigma} = T$$

T is the scalar of the energy tensor of matter.

Secondly Einstein imposed the condition that these equations are covariant with respect to arbitrary transformations of a determinant equal to 1 and replaced (15) by:

$$(17)\ R_{im} = -\kappa \left( T_{im} - \frac{1}{2} g_{im} T \right).$$

Einstein then realized that he could demonstrate that his field equations (17) satisfied the conservation of momentum-energy. He wrote in the final November 25, 1915 paper that, these were the "reasons that motivated me to introduce the second term on the right-hand sides of" (15) and (17). Einstein wrote after equation (17) that $T_{im}$ satisfies the conditions (8) or (12), the energy-momentum conservation.[12] The most interesting point of this result is that, perhaps the term on the right-hand side of equation (8) [or of equation (7) times $g^{\mu\nu}$], which was already written in the November 4, 1915 paper,[13] could give Einstein the idea for the second term on the right-hand side (1/2 $g_{im}$T) of equations (15) and (17). We may conclude that the energy-momentum conservation principle played a crucial role in Einstein's gravitational theory before and in 1915.

---

[10] Einstein, 1915a, p. 783.
[11] See: Weinstein, Galina, "FROM THE BERLIN "ENTWURF" FIELD EQUATIONS TO THE EINSTEIN TENSOR II: November 1915 until March 1916", ArXiv: 1201.5353v1 [physics.hist-ph], 25 January, 2012, pp. 53-56; Einstein, Albert (1915b), "Die Feldgleichungen der Gravitation", *Königlich Preußische Akademie der Wissenschaften* (Berlin). *Sitzungsberichte*, 1915, pp. 844-847; p. 845.
[12] Einstein, 1915b, p. 845; Einstein to Sommerfeld, November 28, 1915, *CPAE*, Vol. 8, Doc. 153.
[13] Einstein, 1915a, p. 782; p. 784.